\documentstyle[aasms4]{article}

\def\eps@scaling{.95}
\def\epsscale#1{\gdef\eps@scaling{#1}}

\def\plotone#1{\centering \leavevmode
    \epsfxsize=\eps@scaling\columnwidth \epsfbox{#1}}

\def\spose#1{\hbox to 0pt{#1\hss}}
\def\simlt{\mathrel{\spose{\lower 3pt\hbox{$\mathchar"218$}}
     \raise 2.0pt\hbox{$\mathchar"13C$}}}
\def\simgt{\mathrel{\spose{\lower 3pt\hbox{$\mathchar"218$}}
     \raise 2.0pt\hbox{$\mathchar"13E$}}}
\def\lsim{\rlap{$<$}{\lower 1.0ex\hbox{$\sim$}}}
\def\gsim{\rlap{$>$}{\lower 1.0ex\hbox{$\sim$}}}

\def\refs{\leftskip=.3truein\parindent=-.3truein}
\def\unrefs{\leftskip=0.0truein\parindent=20pt}

\def \m.        {\rlap{$.$}^{\rm m}}
\def \s.        {\rlap{$.$}^{\rm s}}
\def \am.       {\rlap{$.$}'}
\def \as.       {\rlap{$.$}''}

\def \deg.      {\rlap{$.$}^\circ}

\newcommand{\fcaption}[1]{
        \refstepcounter{figure}
        \setbox\@tempboxa = \hbox{\tenrm Fig.~\thefigure. #1}
        \ifdim \wd\@tempboxa > 6in
           {\begin{center}
        \parbox{6in}{\tenrm\baselineskip=12pt Fig.~\thefigure. #1 }
            \end{center}}
        \else
             {\begin{center}
             {\tenrm Fig.~\thefigure. #1}
              \end{center}}
        \fi}

\newcommand{\tcaption}[1]{
        \refstepcounter{table}
        \setbox\@tempboxa = \hbox{\tenrm Table~\thetable. #1}
        \ifdim \wd\@tempboxa > 6in
           {\begin{center}
        \parbox{6in}{\tenrm\baselineskip=12pt Table~\thetable. #1 }
            \end{center}}
        \else
             {\begin{center}
             {\tenrm Table~\thetable. #1}
              \end{center}}
        \fi}

\begin{document}
\pagenumbering{arabic}

\vspace*{2cm}

\centerline{\bf	The Cosmological Flatness Problem
	and
	Limits on Dark Matter}

\vspace{2cm}

\centerline	{George W. Collins, II}

\centerline	{The Warner and Swasey Observatory
	of
	Case Western Reserve University}
\centerline{10900 Euclid Av., Cleveland, Ohio 44106-7215, USA}

\vspace{2cm}

\centerline {To be published in {\it Comments on Astrophysics} Vol. 18, No. 6.}

\newpage

\centerline{\bf 	Abstract}

	The traditional argument to justify $\Omega_0 = (\rho/\rho_{crit}) =
1$ which suggests an absurdly close agreement between the initial
expansion energy of the universe and its binding energy is reversed to
show that it is unreasonable to expect the current value of W to be
within a factor of ten or more of unity. The Uncertainty Principle is
applied to the initial value of the mass require to close the universe.
The tiny fractional uncertainty is amplified so that the current
fractional uncertainty is of the order of unity. This suggests that the
requirement of exact initial agreement between the binding energy and
the expansion energy is unreasonable as it can never be tested by
contemporary measurement.

\newpage

	Nearly thirty years ago Robert Dicke$^3$ introduced the notion that the
universe must be ``extremely finely tuned" to yield the present
observed balance between the energy of expansion and the gravitational
self binding energy. That balance is usually described in terms of the
ratio of the current local matter density to that required to close the
universe [i.e. $\Omega_0 = (\rho_0/\rho_{crit}) = 1$]. While present
observational limits
on $\Omega_0$ are probably generously described as $0.1 < \Omega_0 < 10$, such
an
apparently wide range suggests a very much narrower range at earlier
epochs. This results basically from the non-linearity of the dynamical
laws governing the expansion. During the vast majority of the time
since the origin of the universe the laws that govern the expansion are
adequately characterized by the those of Newton. Dicke suggests that
the present range of W would have to be decreased to one part in a
thousand at the time galaxies began to form and by 1 part in $10^{13}$
during the age of nuclear reactions when the elements re-formed. A
further reduction is required as one moves back further in time toward
the origin of the universe. If one were to push this argument through
the inflation era to the earliest epoch of the universe one would find
as did Lightman and Gringerich (1992), that $\Omega$ can depart from unity by
only one part in $10^{59}$. The continual approach of W toward unity as one
moves back through the universe through eras of physical structure
having increasing uncertainty has been used by many to assert that
$\Omega_{\rm{i}} =
1$ is a initial condition of the structure of the universe since any
``real" departure from unity as small as 1 part in $10^{59}$ is tasteless.
This argument has become known as the Flatness Problem.

	In this essay, I take an alternate view. I argue that the proper view
of the universe is to follow its evolution from the initial conditions
to the present and what results is what should be expected. However, of
greatest importance is the notion that any observable quantity can have
an exact value. One of the triumphs of the twentieth century is that
the exactness of nineteenth century classical physics has been replaced
with a quantum uncertainty described by Heisenberg's interpretation of
Quantum Mechanics. While such uncertainty in, say, the binding energy
of the universe is likely to be ridiculously small, we have just seen
that ridiculously small departures of $\Omega_{\rm{i}}$ from unity can lead to
substantial departures in the present era. Thus, the very argument used
to force the value of $\Omega_{\rm{i}}$ toward unity should be inverted to show
that
tiny quantum uncertainties in that observable parameter evolve into
significant uncertainties in the present. So we are not allowed to use
the ``Flatness Problem" to force perfection on the initial conditions
of the early universe and then invert the procedure and demand that
$\Omega_0= 1$ in the current era. Indeed, we should be very surprised if
$\Omega_{\rm{i}}$
turned out to be exactly unity. Let us take a moment to see how the
Flatness Problem arose. Then we will attempt to estimate the impact on
the current value of $\Omega_0$ of Heisenberg Uncertainty Principle applied
energy of the universe at the earliest time allowed by contemporary
physics.

	Ever since the interpretation of the large red-shifts of distant
galaxies found by V.M. Slipher (e.g. Osterbrock$^{10}$) as a general
expansion of the universe by Edwin Hubble$^{6,7}$, astronomers have used the
physical laws of the present to describe the  images of the past as
seen in their telescopes. The discovery of radiation from the primeval
fireball by Penzias and Wilson$^{12}$ confirmed an earlier prediction by
Alpher and Herman$^{1,2}$, inspired by Gamow$^{4}$, that was based on
extrapolating the physics of the present far into the early stages of
the past. This venerable approach to understanding the early phases of
the universe continued to push back the era of understanding through a
time when nuclear physics dominated the world (e.g. Weinberg$^{13}$).

	The reversing of the observed expansion of the universe so as to
interpret the early structure of the universe requires that the
earliest phases consist of matter at extreme temperatures and densities
originating in what has come to be called the``Big Bang". Two
fundamental problems with the Big Bang cosmology emerged in the sixties
and were called the Horizon Problem and the Flatness Problem. The
former arises from the similarity of conditions in different parts of
the primeval fireball which could have never been in contact with each
other throughout the million years or so during which they traveled to
the points in space-time where they released the photons observed
today. Such similarity could only be established by mutual interactions
which could never have taken place. When they emitted the photons we
see today they simply existed beyond each other's current event
horizon. Alan Guth (1981) suggested a modification to the standard Big
Bang Cosmology called {\it inflation} wherein an extremely rapid expansion
took place during the earliest phases of the universe which allowed
those widely separated places to
 once have been in contact and thereby to adopt similar characteristics.

	The ``Flatness Problem" is somewhat different in concept. The dynamical
Big Bang Cosmology has three possible outcomes depending on initial
conditions. If the gravitational attraction of the matter in the
universe is strong enough, it will eventually stop the expansion and
cause the universe to collapse to a point in an event sometimes
described as the ``Big Crunch". On the other hand, should the initial
bang have been sufficiently intense then the expansion energy will
overwhelm the self-gravity of the matter and the universe will expand
forever. Clearly there is a special case between these two where the
gravitational self-energy will exactly equal the expansion energy and
the expansion will stop after an arbitrarily large time has elapsed. In
the language of General Relativity these three cases correspond to the
space-time of the  universe being positively curved, negatively curved,
or flat. Simply stated, the ``Flatness Problem" is that an extremely
large time has already passed and the expansion is still going on, but
not too vigorously. This suggests that the universe in which we live is
rather close to being flat. The standard way of describing this is that
\begin{equation}
\Omega \, \equiv  \, \rho/\rho_{cr}\, = \, (8\pi/3){\rm G}\rho/{\rm H}^2 \, =\,
1 ,
\end{equation}
where H is the value of the Hubble constant which describes the rate of
expansion of the universe. Now the value of $\Omega$ changes in time and its
present value can be measured with great difficulty yielding $0.1 < \Omega_0 <
10$ which seems like a horrendous range. But as Robert Dicke (1969)
first pointed out, even this range requires that $\Omega$ must have been much
nearer unity in the past. The problem arises in just how much closer to
unity $\Omega$ must have been.

	While the structure of the early phases of the universe are complicated
by inflation and general relativity, the vast majority of the dynamical
history can be understood in terms of rather classical Newtonian
physics. Following Dicky$^3$ and using simple Newtonian mechanics one gets
\begin{equation}
\frac{1}{2} \rho {\rm v}^2  \, - \,  4\pi {\rm G} \rho^2 \rm{r}^3 / 3  \, = \,
{\rm E}  \, =  \, \Psi + \Phi
\end{equation}
where $\Psi$ and $\Phi$ are the kinetic and gravitational binding energy per
unit volume respectively. E is the total energy density. Since both
$\Psi$ and $\Phi$ were far larger in magnitude in the past while E remains
constant, the relative value of say E$/|\Phi|$ must become smaller and smaller
in the past. One can choose most any epoch in the early history. If one
takes $0.1 < \Omega_0 < 10$ to be representative of the present conditions,
then at the decoupling of matter and radiation at about a million years
$\Omega \gg 1\pm 0.0001$ and $(\rm{E}/|\Phi|) <0.0001$. This represents a
remarkable balance
between the gravitational binding energy and the energy of expansion.
To extend the argument through the earlier era to the first three
minutes when nuclear physics dominated the structure requires that the
agreement between $\Phi$ and $|\Psi |$ must have been good to one part in
$10^{16}$.
Application of inflation pushes the agreement to absurdly high values
ranging between $10^{43}$ to $10^{55}$ in Alan Guth's original exposition of
the
inflation cosmology$^5$. The value is simply determined by when one
chooses to begin the modeling of the universe. The earlier the time,
the closer the agreement between $\Psi$ and $\Phi$ and the nearer
$(\rm{E}/|\Psi |$)
must be to
zero. There is an end to this progression. When one inquires into the
structure of the early  universe, the physical equations fail at a time
known as the Planck time $\rm{t}_{\rm{p}} = 5.4\times10^{-44}$ sec. Lightman
and Gingerich$^8$
place the required agreement at this time to be about 1 part in $10^{59}$.
In other words $(\rm{E}/|\Psi |)) < 10^{-59}$.

	So far there is no real problem for cosmology since the above argument
simply notes the instability in the dynamical equations of motion
whether they arise from Newtonian mechanics, General Relativity, or
during the inflation era. The problem arises when one suggests that a
number so small must really represents an initial condition on the
universe and it is tasteless to suggest that this initial condition is
anything other than $(\rm{E}/|\Psi |) = 0$. If E
was ever zero it is forever zero and
$\Omega_0 = 1$. It is generally considered a triumph of the Inflation Cosmology
that it not only solves the horizon problem, but that by {\it reductio ad
absurdum} it appears to have eliminated the ``Flatness Problem". However,
it is clear that any cosmology that pushed the structure of the
universe back to the Planck time would have reached the same
conclusion.

	So compelling is this argument that it has transcended speculation and
achieved dogma. The real ``Flatness Problem" arises when one places the
statement $\Omega_0 = 1$ in confrontation with the observations. Time and again
careful measurement of the visible matter, and the influence of unseen
baryonic matter on the dynamical behavior of the visible matter, has
let thoughtful observers to conclude that $\Omega_0 \sim 0.3$ and is probably
closer to 0.1. Analysis of the chemical composition of the universe
compared with careful models based on nuclear physics which generate
such a composition also suggested an $\Omega_0 \ll 1$. Additional problems
arise
with the dynamical age of the universe compared with the age of the
oldest stars should $\Omega_0 = 1$. This has led some to suggest a cosmological
constant  and the balance (90\% ) of the matter is comprised of
non-baryonic particles whose existence has yet to be detected. The
consequences of $\Omega_0 = 1$ seem extreme enough that we should re-examine
the foundation of the argument.

	While the quantitative view of the ``Flatness Problem" is usually
characterized by the ratio of the mean density of the universe to the
value required to stop its expansion (i.e. $\Omega$), we note that neither is
a directly observed quantity. Indeed, Peebles$^{11}$ devotes an entire
chapter to methods of determining the current mass density from
observable parameters which are basically the total mass within some
known volume of the universe. There are incredible difficulties in
determining the actual mass of material and one must also be careful in
choosing the volume. Should the volume be too large there will be
systematic variations throughout the volume resulting from the
expansion of the universe itself. If the volume is too small, it will
not contain a representative sample of the mass of the universe and the
result will be biased. Indeed, as we find increasingly large structures
in the universe, it is not obvious that such a volume exists, but will
leave it to others to explore the implications of this possibility.

	Although the quantity  $\Omega$ is usually expressed in terms of relative
densities, it is also equal to the ratio of the observed mass within
the suitable chosen volume to the mass required to stop the expansion
of the material within that volume. Since these are the observed
quantities we shall focus on them in formulating the flatness problem.
This form also makes the balance between the gravitational binding
energy and the expansion energy more transparent. In the case where the
universe is asymptotically balanced, the expansion energy exactly
equals the gravitational binding energy so that the total energy is
zero. Thus a comparison between the actual mass in the carefully chosen
volume and the mass required to stop the expansion is a comparison
between the actual gravitational binding energy and the expansion
energy that would lead to a barely closed universe.

	That we can choose such a volume and ignore the remainder of the
universe can be traced back to Newton who showed that a spherically
symmetric distribution of matter surrounding a spherical volume played
no role in determining the gravitational field within that volume.
There is a parallel theorem known as Birkhoff's Theorem$^9$ that plays the
same role in General Relativity. Thus the expansion dynamics of the
universe can be deduced from the motions and matter within our
specially chosen volume and such an argument can be made at any epoch
in the history of the universe.

	In order to demonstrate the impact of evolving small perturbations or
uncertainties in the initial conditions of the universe to the present
epoch, let us consider the impact of Heisenberg's Uncertainty Principle
on the mass contained within a small volume at the earliest time at
which we may still expect that physics as we know it to describe the
universe -- the Planck time. If that mass is $\rm{m}_{\rm{i}}$ then
\begin{equation}
\delta(\rm{mc}^2)\delta \rm{t} = \hbar
\end{equation}
A reasonable value for $\delta$t is the Planck time itself so that
\begin{equation}
	\delta \rm{m}_{\rm{i}} = \hbar/\rm{t}_{\rm{p}}\rm{c}^2 = 2\rm{m}_{\rm{p}} ,
\end{equation}
where $\rm{m}_{\rm{p}}$ is the Planck mass $(\hbar
\rm{c}/4\rm{G})^{\frac{1}{2}}$.
An alternate way of viewing this uncertainty is to write
\begin{equation}
	\delta \rm{p}\delta \rm{x} = \delta (\rm{m}_{\rm{i}}\rm{c})\delta \rm{x} =
\hbar  ,
\end{equation}
which leads to the same result

\begin{equation}
	\delta \rm{m}_{\rm{i}} = \hbar/\ell_{\rm{p}} \rm{c} = 2\rm{m}_{\rm{p}} ,
\end{equation}
so that the location of m$_{\rm{i}}$ is uncertain by the ultimate smallest
length $\ell_{\rm{p}}$  beyond which we might expect the very equations of
physics to
fail. While this may be an uncomfortably early time for some, it is the
only reasonable time at which to  investigate and apply an initial
condition since it represents the ultimate initial time. Although the
application of initial conditions in general relativity is tricky,
there can be little doubt that some form of quantum theory must apply
to them. Even though the proper reconciliation of quantum theory with
general relativity has yet to be formulated, it is not unreasonable to
expect a fundamental property of the universe expressed by the
Heisenberg Uncertainty Principle to hold as one approaches the point
where those initial conditions apply. If it does not, then the very
concept of an initial condition is unlikely to have any meaning and all
consequences of an initial condition having a particular value are
suspect.

	If we then take this as the uncertainty in the mass that will evolve
into the mass used to determine the observed value of W$_0$, then the
initial uncertainty in W$_{\rm{i}}$ at the Planck time will be
\begin{equation}
	(\rm{d}\Omega)_{\rm{i}} = 2\rm{m}_{\rm{p}}/\rm{m}_{\rm{0}} >
2\rm{m}_{\rm{p}}/\rm{m}_{\rm{c}} ,
\end{equation}
where $\rm{m}_{\rm{0}}$ is the present mass within the volume used to measure
the
current mass density $\rho_0$. This is because we have taken the mass
$\rm{m}_{\rm{0}}$
within the specially selected volume to be conserved by definition when
shrinking the volume back to the earliest allowed time $\rm{t}_{\rm{p}}$. In
order to
make this as small an uncertainty as possible we take the density to be
the closure density $\rho_{\rm{c}} =(3\rm{H}_0^2/8\pi \rm{G})$
and the corresponding mass to be the
closure mass $\rm{m}_{\rm{c}}$ so that the uncertainty in the initial value
of $\Omega_{\rm{i}}$
becomes
\begin{equation}
	(\delta\Omega)_{\rm{i}} > 2\rm{m}_{\rm{p}}/(4\pi \rm{R}_0^3 \rho_{\rm{c}} /3)
= 2\rm{Gm}_{\rm{p}}/(3\rm{H}_0^2 \rm{R}_0^3) .
\end{equation}
Here $\rm{H}_0$ is the current value for the Hubble constant. If we take the
radius of the volume used to determine the current mass density and the
associated closure density to be 1000 Mpc then
\begin{equation}
	(\delta\Omega)_{\rm{i}}\geq  4.71\times 10^{-60}/\rm{h}^2 ,
\end{equation}
where h is the current Hubble constant in units of 100 km sec$^{-1}$
Mpc$^{-1}$. Thus
we see if $\rm{h} < 1$, the uncertainty in $(\delta\Omega)_{\rm{i}}$ is
likely to be of the order
of $10^{-59}$. However, that is just the inverse of the amplification factor
(e.g. Lightman and Gingerich$^8$) which $\delta\Omega$
will experience in evolving to
the present. Thus we could expect the initial indeterminacy of the
initial value of $\Omega_i$ to evolve into an uncertainty in $\Omega_0$ of
\begin{equation}
	(\delta\Omega)_0 \cong 1.
\end{equation}

	It is worth noting that choosing R$_0$ in equation (8) to represent a
smaller volume increases $(\delta\Omega)_{\rm{i}}$, which
weakens the flatness constraint
even further. The same result applies for using a smaller matter
density such as the very uncertain observed value of the matter density
$\rho_0$. If one goes to the other extreme and chooses the largest possible
volume for determining $\rho_0$, then one could take
$(\rm{R}_0 = \rm{c}/\rm{H}_0)$ to be the
size of the visible universe and one gets that

\begin{equation}
	(\delta\Omega)_{\rm{i}}  >  \rm{t}_{\rm{p}}\rm{H}_0 \cong  10^{-60} ,
\end{equation}
which suggests that
\begin{equation}
	(\delta\Omega)_0  >  0.1 .
\end{equation}
However, choosing such a large volume to determine the mean density
would include systematic and model dependent corrections introducing
far larger uncertainties in $\Omega_0$ than those resulting from the amplified
uncertainty of the initial conditions.

	Thus it would seem that if the foundations of quantum mechanics are
applicable during the very earliest phases of the universe, initial
conditions related to present day observables should have tiny
uncertainties which have evolved into significant uncertainties today.
This does not mean that the contemporary value of $\Omega_0$ cannot be measured
to greater precision than that suggested by the value of $\delta \Omega_0$.
It does
suggest that such a measured value cannot place limits on the initial
conditions of the universe that exceed those estimated by equation
(10). The thrust of this argument is that it is unreasonable to expect
any initial value of a dynamical constraint on the evolution of the
universe to be known with absolute precision and that plausible
uncertainties in that initial value lead to results which are today
totally compatible with observation. Thus one cannot use the so-called
``Flatness Problem" to justify the existence of non-baryonic matter or
non-zero values of the cosmological constant since such just ifications
require the absolute equality of the expansion energy and the
gravitational binding energy (i.e. $\Omega_0 = 1.00... \pm 0.00...$) .

	While a value of $\Omega_0$ on the order of a few tenths must still be
regarded
as remarkable, it is not magical or mystical in any way. The observed
value also has the benefit of reducing some of the current difficulties
in reconciling the dynamical age of the universe with the age of its
oldest constituents. One can still have the philosophically satisfying
result that the total energy of the universe is as close to zero as one
could expect it to be without being forced to the extreme conclusions
compelled by it being identically zero. The argument simply says that
the universe is as well behaved as quantum mechanics will allow it to
be. But perfection at any era is unattainable, and conclusions which
require perfection should be regarded as suspect.

	Since the mass of the observable universe increases steadily in time,
it is a fair question to ask if it will ever become large enough for
the quantum uncertainty in $\Omega_{\rm{i}}$ to be small enough to allow for a
measurement of $\Omega_0$ which would place stringent enough limits on the
initial conditions to justify perfection.  From the expression for
$(\delta \Omega)_{\rm{i}}$
obtained for the largest observable mass that could be used for
obtaining its minimum value, it is clear that the smallest expected
value decreases linearly with the age of the universe. However, as
Dicke$^3$ shows, the expansion of a zero pressure universe characteristic
of its late stages grows as t$^{2/3}$. Thus the gravitational binding energy
will approach zero as t$^{-2/3}$ implying the amplification factor of the
initial uncertainty will continue to grow as t$^{2/3}$. This would suggest
that if one could use the entire mass of the visible universe to
determine the actual mean density of the universe, its difference from
the value required to exactly close the universe would approach zero
as the cube root of the Hubble age of the universe. This would appear
to hold out some hope for those classicist who seek perfection between
the expansion energy and the gravitational binding energy. Since an
asymptotically flat universe contains an arbitrary amount of matter,
the time will come when the mass of the visible universe becomes
arbitrarily large. Since the uncertainty in the initial condition for
the observable universe remains the Planck mass, the initial
uncertainty will become arbitrarily small. Fortunately the dynamical
growth of the uncertainty in the observable mass grows slower than the
initial value decreases so the idyllic result of an exact balance in
the initial condition for the entire observable universe that can
eventually exist can be realized. The unfortunate aspect of this
argument is that a measurement which could have convincing accuracy
lies a thousand times the current Hubble age in the future.

\noindent{\bf References}

\refs

1.      Alpher, R.A. and Herman, R.     Nature, 162, pp. 774, (1948).

2.      Alpher, R.A. and Herman, R.     In {\it Modern Cosmology in
Retrospective}, (ed. B.Bertotti, R.Balbinot, S. Bergia, and A. Messina)
Cambridge University Press, Cambridge, pp.137-140, (1990).

3.      Dicke, R.H.                     {\it Gravitation and the Universe},
Jayne
Lectures for 1969, American Philosophical Society, Independence Square,
Philadelphia, pp. 55-60, (1970).

4.      Gamow, G.                       Nature, 162, pp 680, (1948).

5.      Guth, A.                        Phys. Rev. D., 23, pp. 347-356,
(1981).

6.      Hubble, E.                      Proc. of the Natl. Acad. Sci.
15, pp. 168, (1929),

7.      Hubble, E.                      {\it The Realm of the Nebula}, Yale
University Press, New Haven, pp. 102- 123, (1936).

8.      Lightman, A., and Gingerich, O. Science 255, pp.690-691,
(1992).

9.      Misner, C. W., Thorne, K.S., and Wheeler, J.A.,
					{\it Gravitation}, W.H.Freeman, San
					Francisco, pp. 843-844,
					(1973).

10.     Osterbrock, D.O.                In {\it Modern Cosmology in
Retrospective}, (ed. B.Bertotti, R.Balbinot, S. Bergia, and A. Messina)
Cambridge University Press, Cambridge, pp. 258-275, (1990).

11.     Peebles, P.J.E.                 {\it Physical Cosmology}, Princeton
University Press, Princeton N.J. pp. 56-115, (1971).

12.     Penzias, A.A., and Wilson, R.W. Ap.J. 142, pp. 419, (1965).

13.     Weinberg, S.                    {\it The First Three Minutes}, Basic
Books, Inc. New York, (1977).

\unrefs

\end{document}